# Microwave power and chamber pressure studies for single-crystalline diamond film growth using microwave plasma CVD

Truong Thi Hien[1], Jaesung Park[2], Cuong Manh Nguyen[3], Jeong Hyun Shim[1,4,*], Sangwon Oh[3,†]

[1] Quantum Magnetic Sensing Group, Korea Research Institute of Standards and Science, Daejeon 34113, South Korea

[2]Tactile Standard Convergence Research Team, Korea Research Institute of Standards and Science, Daejeon 34113, South Korea

[3]Department of Physics, Ajou University, Gyeonggi-do, 16499, South Korea

[4]Department of Applied Measurement Science, University of Science and Technology, Daejeon 34113, South Korea

Corresponding Author Email address: *jhshim@kriss.re.kr / †sangwonoh@ajou.ac.kr

## Abstract

Single-crystalline diamond (SCD) films possess exceptional thermal, chemical, and optical properties, making them ideal for advanced applications. However, achieving uniform film quality via microwave plasma chemical vapor deposition (MPCVD) remains challenging due to spatial variations in plasma characteristics. This study systematically examines the influence of microwave power and chamber pressure on the growth of SCD films using $CH_4/H_2$ gas mixtures. Under optimized conditions (3,900 W, 120 Torr), the films exhibit low surface roughness (~2.0 nm), a sharp $sp^3$ Raman peak at 1,332.2 $cm^{-1}$, and no detectable C–H related features, indicating high crystalline purity. Cross-sectional TEM analysis confirms a uniform (100)-oriented single-crystal structure across the entire sample. These findings advance the understanding of the interplay between deposition parameters and film quality, and establish a more robust foundation for optimizing MPCVD processes in large-area, high-purity diamond fabrication.

## 1. Introduction

Diamond possesses exceptional physical and chemical properties, including a wide bandgap, high carrier mobility, superior thermal conductivity, infrared transparency, and excellent chemical stability [1,2]. These properties make diamond ideal for diverse applications, where smooth, uniform films are crucial, especially in optics, electronics, and biomedicine. However, impurities in natural and HPHT diamonds limit their usability [3]. In contrast, chemical vapor deposition (CVD) has emerged as a versatile method for synthetic diamond growth, employing techniques such as HF-CVD, DC/arc plasma jets, DC-glow discharge, PCVD, and MP-CVD. Among these, MP-CVD is most effective for producing high-purity single-crystal diamond (SCD) films with atomically flat surfaces. However, this method often suffers from center-to-edge variation in film quality, attributed to non-uniform microwave distribution, plasma shape asymmetry, and gradients in temperature and electric field across the substrate [4,5].

Various research strategies have been employed to address this issue, including the designs of chamber shapes and molybdenum substrate holders to enhance the uniformities of microwaves, electric fields, and temperature over the sample surface [6,7]. In recent years, advancements in MP-CVD reactor design and in situ diagnostics have provided deeper insights into growth uniformity. Optical emission spectroscopy (OES) and cavity ring-down spectroscopy (CRDS) have enabled real-time monitoring of reactive species and plasma dynamics [4,5,8]. Additionally, studies on microwave power, chamber pressure, and gas composition have helped define optimal conditions that balance diamond deposition with etching of non-diamond phases, ensuring surface-localized nucleation and steady carbon incorporation into the lattice. Furthermore, these strategies minimize variations in the distribution of the $CH^*_3$/H radical ratio, electric field, and temperature across the sample surface from center to edge, thus improving the uniformity of the diamond film. In particular, the $CH^*_3$ and H intermediate radicals produced from the decomposition of $CH_4$ and $H_2$ play a crucial role. The interaction of these radicals with the substrate initiates a series of reactions that facilitate the incorporation of carbon atoms into the diamond lattice, promoting the growth of the diamond film [5,9–12]. Teraji et al. reported successful control of microwave power ranging from 3,800 to 4,200 W and variation of $CH_4$ concentrations from 0.4 to 32.0%, resulting in high-quality diamond films [13]. Vikharev et al. demonstrated that, under optimized microwave plasma CVD growth conditions, high-quality single-crystalline diamond films were obtained, devoid of non-epitaxial diamond crystallites, hillocks, or a polycrystalline diamond rim around the top surface [14]. The typical pressures for high-quality diamond film growth regimes in methane-hydrogen plasma are reported within 100–300 Torr [8,15,16]. Yang et al. have illustrated the non-uniform distribution of $CH^*_3$/H ratio, electric field, and temperature across the sample, resulting in disparate qualities of diamond film from the center to the periphery [17].

Furthermore, the variation in the $CH^{*}_3/H$ ratio under different pressures, as noted by Ashfold et al., has been acknowledged to influence the deposition of diamonds [18]. Diamond films deposited using the MPCVD method typically exhibit uniform nucleation and growth mechanisms. High-quality films are achieved when carbon atoms are consistently and continuously integrated into the crystal lattice, resulting in a very smooth diamond film surface with minimal defects. Therefore, the quality of the diamond film is closely related to its surface roughness [12,19]. However, comprehensive research investigating the influences of both microwave power and chamber pressure on high-quality diamond film with atomic-level roughness, as well as the uniformity of diamond film surface at both the sample center and sample edge, has not been previously reported.

In this study, the MP-CVD deposition of diamond films was investigated using a $CH_4$ and $H_2$ mixture with a small addition of $O_2$ gas, under various conditions. Microwave power and chamber pressure were systematically varied to assess their impact on morphology uniformity, surface roughness, and crystalline quality at the center and edge of the samples. Observed trends in diamond film quality and uniformity are thoroughly analyzed and explained, offering a clearer understanding of the diamond growth process via the MP-CVD method.

## 2. Experiment

The diamond films were grown on (100)-oriented HPHT diamond substrates (from Elements Six Ltd.), $3 \times 3 \times 0.25$ mm$^3$ (semi-open type), using an MP-CVD system (Seki Diamond SDS6350, Cornes Technologies Ltd.). The HPHT diamond substrates were acid-cleaned under similar conditions of mixed acids $HNO_3$:$H_2SO_4$:$HClO_4$ = 1:1:1 in volume ratio at 250 ºC for 1 hour, before and after the growth, and the root mean square roughness (RMS) before the growth was approximately 1.96 nm over a $10 \times 10$ μm$^2$ area. An outer diameter of 50 mm and a thickness of 3 mm Mo holder were used to hold the substrates. The holder had a $3.1 \times 3.1 \times 0.1$ mm$^3$ recess, which held the substrate during the cooling down of the deposition process. Before gas introduction, the chamber was evacuated to a base pressure below $7 \times 10^{-8}$ Torr to eliminate residual contaminants. Ultra-high purity $H_2$, $O_2$, and $CH_4$ gases (≥99.999%) were then introduced at controlled flow rates of 490, 1, and 10 sccm, respectively. A water-cooled substrate holder was employed to maintain thermal stability during plasma operation. The existence of $O_2$ is known to improve diamond film quality [20]. The growth time of the samples was in the range of 4.8 to 5.5 h. The temperature at the surface of the substrates was monitored using a pyrometer (PRO2A, Williamson IR) through a quartz window at the top of the chamber. The Plasma-assisted CVD diamond growth process using $CH_4$ and $H_2$ as input gases is shown in Fig. 1. The morphology of the diamond films was examined using optical microscopy (OM, DM4M, Leica microsystems), field emission scanning electron microscopy (FE-SEM, FEI-Sirion 400 NC, Philips), and atomic force microscopy (AFM, XE7, Park system). The films were characterized using Raman spectroscopy with an excitation wavelength of 532 nm and surface mode (WITec alpha300, Oxford Instruments), and secondary ion mass spectrometry (SIMS, IMS 7f, CAMECA). The SIMS was used to confirm the deposition rate and concentration of impurities such as N. Additionally, cross-sectional TEM samples with a thickness of approximately 70 nm were prepared along the (100) orientation using the standard TEM preparation technique for single crystals, employing Ga ion milling followed by Ar final cleaning with a scanning electron microscope (NX5000, Hitachi). TEM analysis was conducted using a Tecnai TF20 G2 (Thermo Fisher Scientific) transmission electron microscope operated at 200 kV. Selected area electron diffraction (SAED) patterns were acquired using an aperture with a diameter of 750 nm in the specimen plane. High-resolution TEM images were obtained in bright-field mode.

## 3. Results and discussion

### 3.1. Effect of Microwave Power

To understand the role of microwave power on the grown diamond film, we varied the microwave power from 2,700 W to 4,300 W and analyzed the surface using OM, SEM, AFM, and Raman spectroscopy. The OM images at different microwave power and the average temperature during the deposition are shown in Fig. 2a. The average temperature on the sample surface increases with the increase of microwave power. The Mo holder was water-cooled to maintain a stable substrate temperature during diamond film deposition [21]. However, the average sample surface temperature tended to increase with the rise in microwave power. This is because the increased plasma power leads to a higher plasma density, which enhances heat transfer to the sample surface. Additionally, the increased ion emission resulting from the higher microwave power contributes to the heating of the sample surface [22]. Noticeable differences in brightness were observed with lower microwave powers, ranging from 2,700W to 3,300W. These variations could be due to the surface imperfections and the increased grain boundaries, which affect brightness distribution within the images. Increasing the microwave power to 3,900W, resulted in the most uniform distribution brightness distribution. However, with microwave powers exceeding 3,900W, the sample's edge region gradually becomes darker. This inhomogeneity in the sample could be explained by the non-uniform distribution of radical density, electric field, and surface temperature between the sample's center and edge, as demonstrated in relevant studies [9,17,23].

To get a better understanding of the inhomogeneous brightness distribution, the morphological analysis of both the central and peripheral regions of the grown films was performed as depicted in Fig. S1 and Fig.S2. The SEM images were converted to binary form with black background and white pits and analyzed by ImageJ software to obtain the results of pit, particle density and average pit, particle size as shown in Fig. 3a and 3b, respectively. The diagrams these figures illustrate a reduction in both pit density and pit size at the center and edge of the sample as the applied power increases. These observations could explained by the findings of Butler et al.; the growth of diamond films is influenced by the carbon incorporation into the lattice at activated sites, and the density of the activated sites increases with temperature [12]. Furthermore, a decrease in the $CH^{*}_{3}/H$ ratio may result in an etching rate that exceeds the deposition rate, thereby increasing the density of pits, similar to previous reports [15,24]. Hence, the lower microwave power could lead to lower temperatures on the substrate surface, as shown in Fig. 2 (a), resulting in denser pits on the surface. However, the surface morphologies at the edges of the films show a different trend, with larger crystalline structures appearing as dominant surface features. This radial disparity may arise from geometric effects related to reactor design, such as the use of a thin Mo holder, which can influence local plasma confinement and radical transport. Simulation results from previous studies using comparable MPCVD configurations have demonstrated spatial variations in reactive species concentration across the substrate [17]. Although our work did not directly measure this gradient, the consistent morphological difference observed supports the presence of a radially non-uniform growth environment. The increased $CH^{*}_{3}/H$ ratio in the edges of the substrate increased the deposition rate over the etching by H, resulting in the lower pit density under the low microwave powers and larger crystalline structure under the high microwave powers in the edge of the films [15,20,23,25]. In our experiments, a smooth surface was obtained both at the center and the edge of the samples deposited at a microwave power of 3,900 W, with an average substrate temperature of 870°C. This can be attributed to the balance achieved between the density of activated sites and the radially uneven $CH^{*}_{3}/H$ ratio across the HPHT substrates.

The surface roughness of the diamond films at each microwave power was measured using AFM. AFM images of a $10 \times 10$ μm² scan area were captured at both the center and edge of the films, as shown in Fig. S3 and S4. The relationship between microwave power and surface roughness at the center and edge of the samples is presented in the diagram in Fig. 3c. The RMS values of the surface roughness at the center and edge of the film decrease as the microwave power increases up to 3,900W, and the roughness at the sample center remains unchanged with further increases in the power up to 4,300W. The smallest RMS at the center and edge of the films was 2.0 nm when the microwave power was 3,900W. The smallest RMS values are comparable to the RMS of the bare HPHT substrate. Tỏ

The Raman spectra at the center and edge of the grown diamond films from various microwave powers are shown in Fig. 6. We can find three different peaks in the spectra: the diamond phase ($sp^3$) peak at around 1,332.2 $cm^{-1}$, C-H bending at 1,416 $cm^{-1}$, and C-H stretching at 3,119 $cm^{-1}$ [26–29]. The Raman spectra of the microwave power of 3900W show a strong peak at 1,332.2 $cm^{-1}$ and are almost free of the other peaks, indicating that the grown diamond could be comparable to the HPHT substrate. This is a clear sign of a high-quality single-crystal diamond (SCD) [30,31]. However, Raman spectra in the other conditions have C-H bending and stretching-related peaks. This could be explained by the higher etching rate at lower microwave power due to the decrease of the activated centers and enhanced $CH^{*}_{3}/H$ ratio at higher microwave power [17,25,32].

The diamond peaks at the center and edge of samples grown in different microwave powers are expanded for clear comparison in Fig. 7. The Raman spectra of samples were in the 1331,8 to 1332,6 $cm^{-1}$ range, with an insignificant shift within 1 $cm^{-1}$. The full width at half maximum (FWHM) of the peaks from the microwave powers are compared in Fig. 7 c. The FWHM decreases as the microwave power increases up to 3,900W and marginally decreases at the higher microwave powers. The thickness and growth rate of diamond films were found from SIMS measurements, which were performed only at the center of the films. The growth rate of diamond films was found from SIMS measurements, which were performed only at the center of the films. It is widely observed that an increase in temperature enhances the $CH^*_3/H$ concentration ratio, thereby leading to higher deposition rates [22,33,34]. However, some studies have also reported a reduction in growth rate under excessive power and pressure conditions. This behavior may be attributed to plasma imbalances at elevated temperature, which promotes gas-phase crystal formation and reduces the deposition rates [11,35,36]. In our experiments, we observed a growth rate of approximately 1.2 μm/h, which slightly varies as the power increases. This finding can be attributed to the plasma imbalance as well as the type of Mo holder employed [6,16,17,37,38]. However, to our knowledge, this behavior has not been previously reported with Mo holder structures similar to those used in our study. Therefore, a study on the effect of semi-open rack type on deposition rate was required to clarify this further.

### 3.2. Effect of chamber pressure

We also investigated the effect of chamber pressure on diamond films, such as morphology and composition. We fixed the microwave power at 3,900W and varied the pressure from 100 to 140 Torr by utilizing pressure sensors and feedback systems while keeping the gas flow rates the same: $490H_2/1O_2/10CH_4$. The temperature at the substrates stayed around 870 ºC, as shown in Fig. 8a. The inhomogeneous brightness distribution occurred when the pressure was higher than 130 Torr, Fig. 8b-f.

To investigate the uneven distribution of brightness, the grown diamond films were analyzed using SEM, with the results presented in Fig. S5 and Fig. S6. The SEM images were converted to binary form with black background and white pits and analyzed by ImageJ software to obtain the results of pit, particle density and average pit, particle size as shown in Fig. 3a and 3b, respectively. The presence of large crystal grains on the sample surface under low-pressure conditions is consistent with previous reports regarding diamond particle formation in the gas phase at low pressure [39]. The pit density at both the center and the edge of the films increases as the pressure exceeds 130 Torr aligns with the findings of Ashfold et al. This can be attributed to the reduced $CH^*_3/H$ ratio at elevated pressures ranging from 100 to  140 Torr [18]. The increased pit density at the center of the sample may be explained by the higher $CH^*_3/H$ ratio at the edge of the film, as proposed by Yang et al [17].

The RMS roughness of the diamond films at each pressure was measured using AFM. The AFM images with a scan area of $10 \times 10\ \mu m^2$ were taken at the center and edge of the films as the microwave cases, and shown in Fig. S7 (center) and Fig. S8 (edge). The relationship between microwave power and surface roughness at the center and edge of the samples is presented in the diagram in Fig. 7c. The RMS values of the surface roughness suddenly deteriorate as the pressure becomes higher than 120 Torr, where the minimum RMS value of 2.0 nm was obtained at 120 Torr.  This could be explained by the reduced $CH^*_3/H$ ratio at the higher pressure as Ashfold et al. suggested [18].

The Raman spectra at the center and edge of the grown diamond films from various pressure conditions are depicted in Fig. 8. The singular peak at 1,332.2 $cm^{-1}$ in the case of 120 Torr confirms its high quality as the HPHT substrate [10]. However, Raman spectra from the other films revealed a C-H stretching defect peak at 3,119 $cm^{-1}$ on both the central and edge areas [40]. Jiang et al. reported that defects formed at large crystals at low pressure enhanced C-H defect intensity [41]. Additionally, the higher pressure than the optimal pressure of 120 Torr would enhance the $H/CH^*_3$ ratio, leading to increased surface termination by hydrogen atoms and intensifying the persistent vibration [42]

Fig. 9a and 9b illustrate the FWHM of Raman spectra at the central and edge surfaces of the samples, respectively. The Raman spectra of samples deposited in different chamber pressures were in the 1331,7 to 1332,6 $cm^{-1}$ range, appearing the negligible shift within 1 $cm^{-1}$. The FWHM of 7.86 $cm^{-1}$ at the lower pressure is comparable to that of the HPHT substrate, whereas the FWHM at the higher pressure than 120 Torr becomes broad due to the defects in the diamond film, as seen in SEM and AFM measurements. The highest deposition rate of 1.3 μm/h was obtained at 120 Torr.-This agrees with the previous studies that have demonstrated that the deposition rate initially increases with pressure but subsequently decreases once the pressure surpasses the region of optimal plasma equilibrium conditions [11,36].

### 3.3. Discussion.

Fig. 10 and Fig. 11 displays the SAED patterns obtained from both the center and edge of the diamond film grown at 3900 W and 120 Torr exhibit sharp, four-fold symmetry with interplanar spacings of 0.172 nm, 0.121 nm, and 0.086 nm calculated from the SAED patterns corresponding to the (200), (220), and (400) planes, respectively - confirming an FCC diamond structure oriented along the [100] zone axis [43–45]. HRTEM images from both regions reveal well-defined (400) lattice fringes with d-spacings of ~0.086–0.087 nm. The corresponding FFTs from RHTEM images show symmetric, sharp diffraction spots without diffuse scattering, diffraction spots 1, 2, 3 as marked in the FFT images, and the central diffraction spot forming a square indicating high single-crystal quality of diamond film [44,46]. EDS mapping of the entire FIB-prepared TEM sample revealed only carbon and oxygen, with consistent atomic concentrations at both the center and edge, similar to about 99.73% C and 0.27% O, respectively. The low oxygen content, primarily located in the protective Pt layer used during TEM preparation, indicates negligible formation of secondary phases. This is consistent with Raman spectroscopy, which showed no evidence of C–O stretching modes around 1100–1200 $cm^{-1}$.

This study investigated the influence of microwave power and chamber pressure on the morphology, roughness, crystallinity, and growth rate of single-crystalline diamond (SCD) films deposited on HPHT substrates using microwave plasma CVD. Despite being studied separately, these two parameters are fundamentally interrelated in determining the radical species balance, surface kinetics, and gas-phase behavior.

As microwave power increases, plasma density and substrate temperature rise, resulting in an enhanced generation of $CH_3$ radicals and atomic hydrogen. This shift modifies the $CH^*_3$/H ratio, which strongly governs the competitive processes of surface etching and diamond nucleation. Under low microwave powers or high pressures, a reduced density of activated surface sites and/or a lower $CH^*_3$/H ratio results in the dominance of etching or non-diamond phase formation, as observed from increased pit density and broad Raman peaks. In contrast, at an optimal power of 3,900 W and 120 Torr, the balance between radical species flux and surface activation supports a growth regime favoring epitaxial diamond formation with minimal defects. This is evidenced by low surface roughness, high-quality Raman signals, and FFT-confirmed crystallinity in HRTEM images.

Moreover, the non-uniformity observed between the film center and edge reflects spatial variation in plasma distribution and gas composition across the substrate surface, consistent with simulations reported by Yang et al. [17]. Based on the observed differences in surface morphology between the center and edge of the samples, we hypothesize a radial variation in $CH^*_3$/H ratio, which has been theoretically predicted in similar MPCVD systems using a thin Mo holder [17]. While direct measurements of radical distributions were beyond the scope of this study, the trends in pit density and crystalline structure are consistent with this interpretation. Future studies employing in situ optical emission spectroscopy (OES) or cavity ring-down spectroscopy could validate this mechanism quantitatively.

The results suggest a growth mechanism controlled by the interplay between three key factors: (1) the $CH_3$/H ratio, (2) density of activated surface sites, and (3) suppression of gas-phase particle formation. These findings support a working hypothesis that optimal single-crystal growth occurs within a narrow window of plasma and pressure conditions that maintain high surface reactivity while preventing secondary nucleation pathways. Further investigation with in situ plasma diagnostics and advanced spectroscopic analysis (e.g., STEM-EELS or TOF-SIMS) is required to refine this mechanism at the atomic scale.

## 4. Summary

In this work, we systematically investigated the effects of microwave power and chamber pressure on the growth of single-crystalline diamond films using microwave plasma CVD. Our results demonstrate that diamond film quality is highly sensitive to both parameters, with optimal conditions observed at 3,900 W and 120 Torr. Under these conditions, the films exhibited minimal surface roughness (~2.0 nm), high structural uniformity, and Raman features comparable to the HPHT substrate, confirming the high crystallinity of the grown SCD films as evidenced by a sharp $sp^3$ peak at 1,332.2 $cm^{-1}$ and the absence of C-H related peaks at 1,416 and 3,119 $cm^{-1}$ under optimal growth conditions. TEM analysis further confirmed the presence of a high-quality, (100)-oriented, face-centered cubic single-crystal diamond lattice structure at both the center and edges of the sample.

The study reveals that achieving high-quality single-crystal growth requires a careful balance between $CH_3$ and H radical flux, substrate surface activation, and gas-phase reaction suppression. Morphological variations between the center and edge regions highlight the importance of reactor configuration and plasma uniformity, especially in scale-up applications. These findings contribute to a deeper understanding of the process–structure relationship in MPCVD diamond growth and provide a predictive framework for optimizing growth parameters in future high-purity, large-area SCD fabrication.

**Declaration of competing interest**

The authors declare that they possess no discernible competing financial interests or personal affiliations that might have conceivably influenced the research findings presented in this paper.

**Acknowledgments**

This research was supported by a grant (GP2024-0013) from Korea Research Institute of Standards and Science, the Institute of Information & communications Technology Planning & Evaluation (IITP) grant funded by the Korea government (MSIT)(2022-0-01026, RS-2023-00230717, No. 2021-0-00076), and a grant funded by the Ministry of Education (No. RS-2023-00285390). We appreciate advice from Dr. Kwak Taeyeong at Tech University of Korea.

# Figures

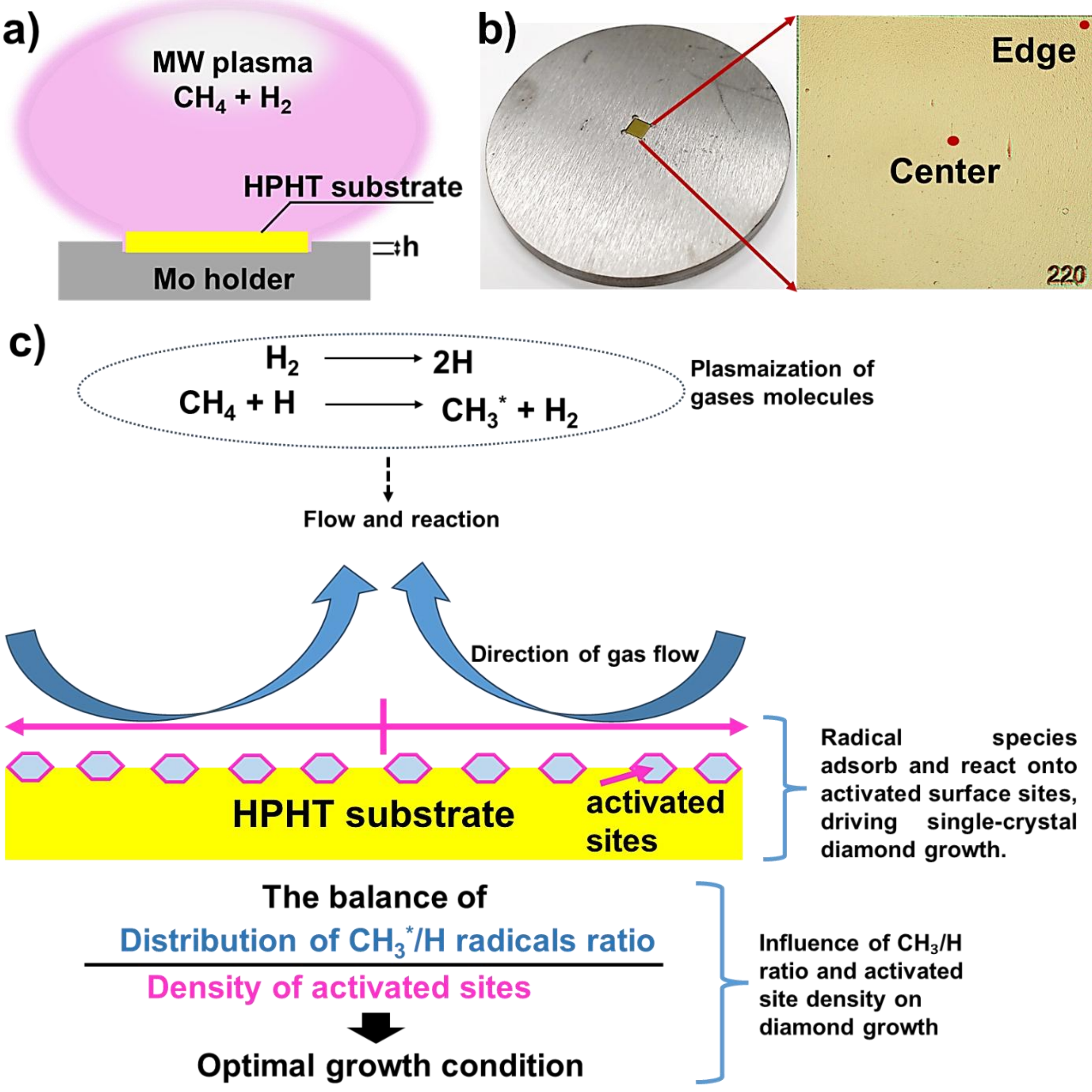

**Fig. 1.** Plasma-assisted CVD diamond growth process using $CH_4$ and $H_2$ gases. a) The illustration of the geometry of the Mo substrate holder. b) The image of the HPHT diamond plate on the Mo substrate holder and the marked surface of the diamond sample indicates the positions used for analysis at the center and edge of the sample. c) Schematic showing the interaction and deposition of diamond via Plasma - CVD method.

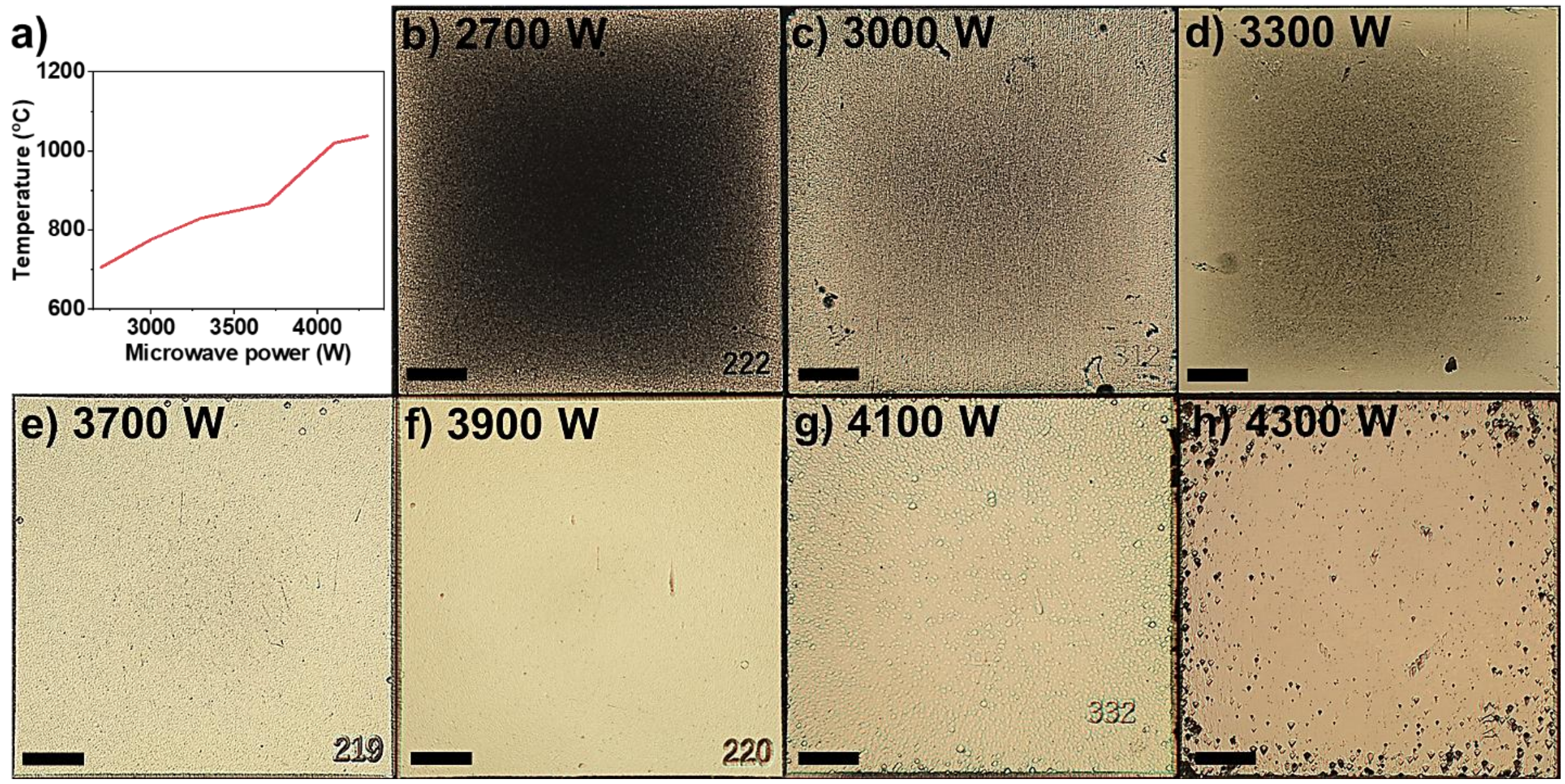

**Fig. 2.** (a) Diagram illustrating the relationship between the sample surface temperature and microwave power. (b-h) Optical images of the samples deposited in different microwave power conditions ranging from 2,700 W to 4,300 W, respectively at 2.5x magnification. The scale in all images is 500 μm.

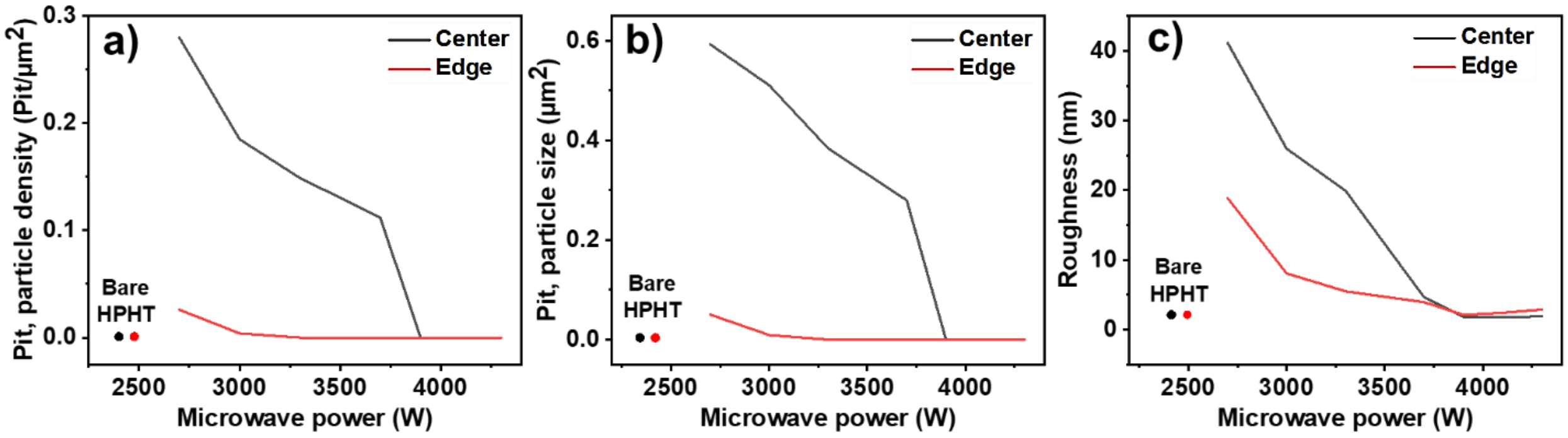

**Fig. 3.** The diagrams illustrate the relationships between a) the pit density, b) average pit size, and c) surface roughness at both the center and edge regions of the samples with the microwave power.

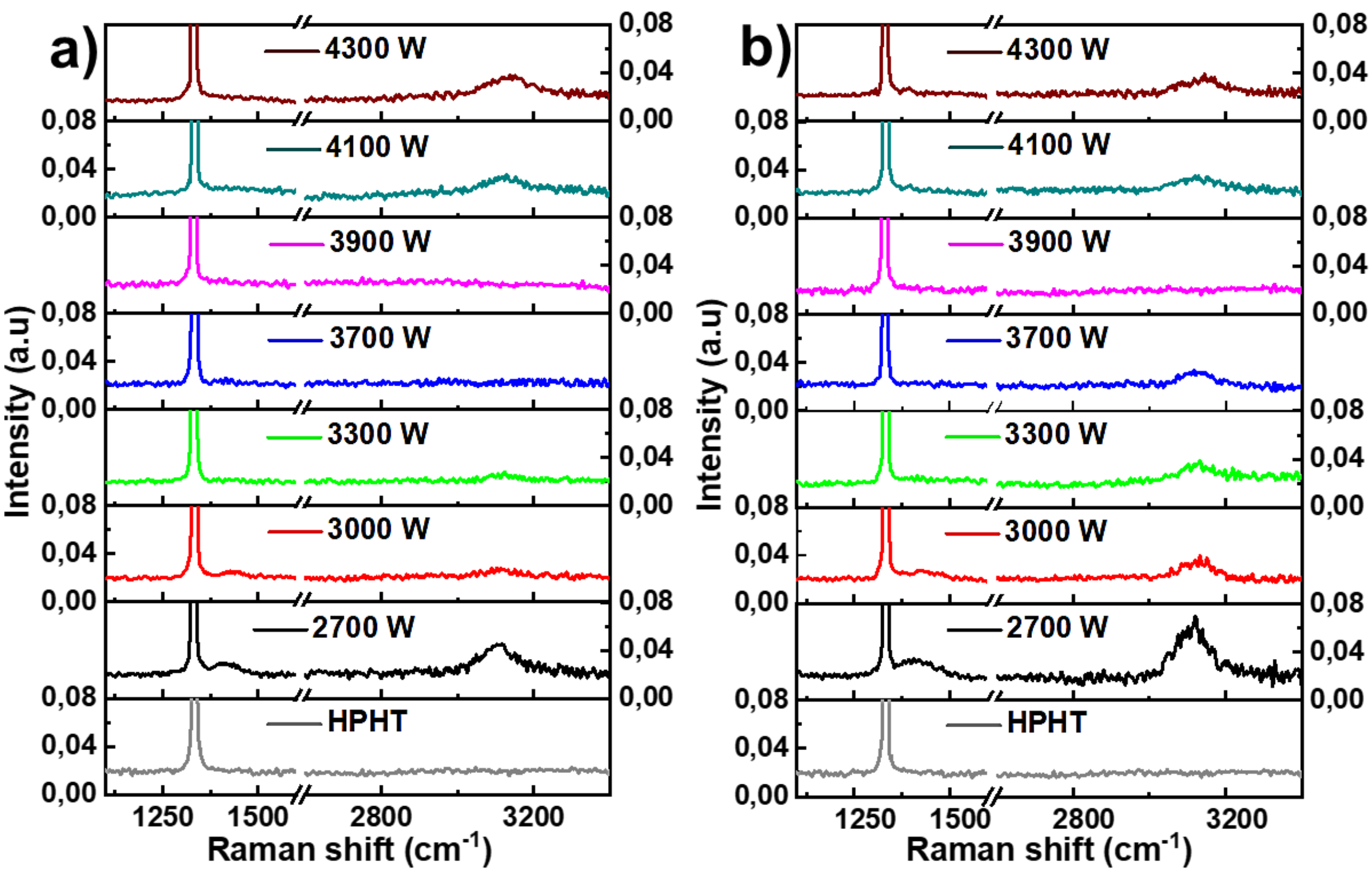

**Fig. 4.** Raman spectra at (a) center and (b) edge of samples deposited in different conditions of microwave power ranging from 2,700 W to 4,300 W.

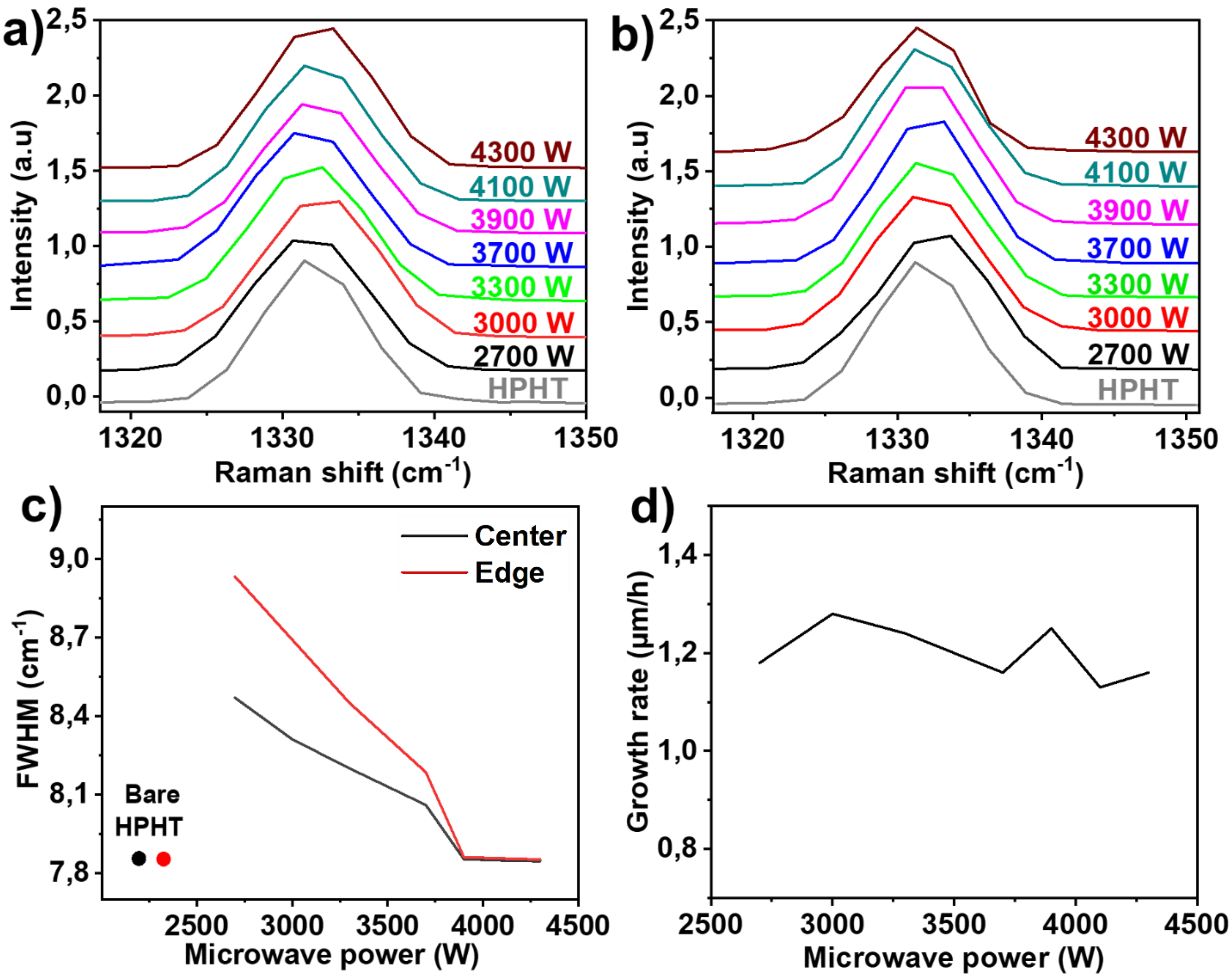

**Fig. 5.** The single diamond peak at (a) the center and (b) the edge of samples deposited in different conditions of microwave power, (c) the comparison of FWHM of samples, and (d) the diagram reveals the relationship between the growth rate and microwave power.

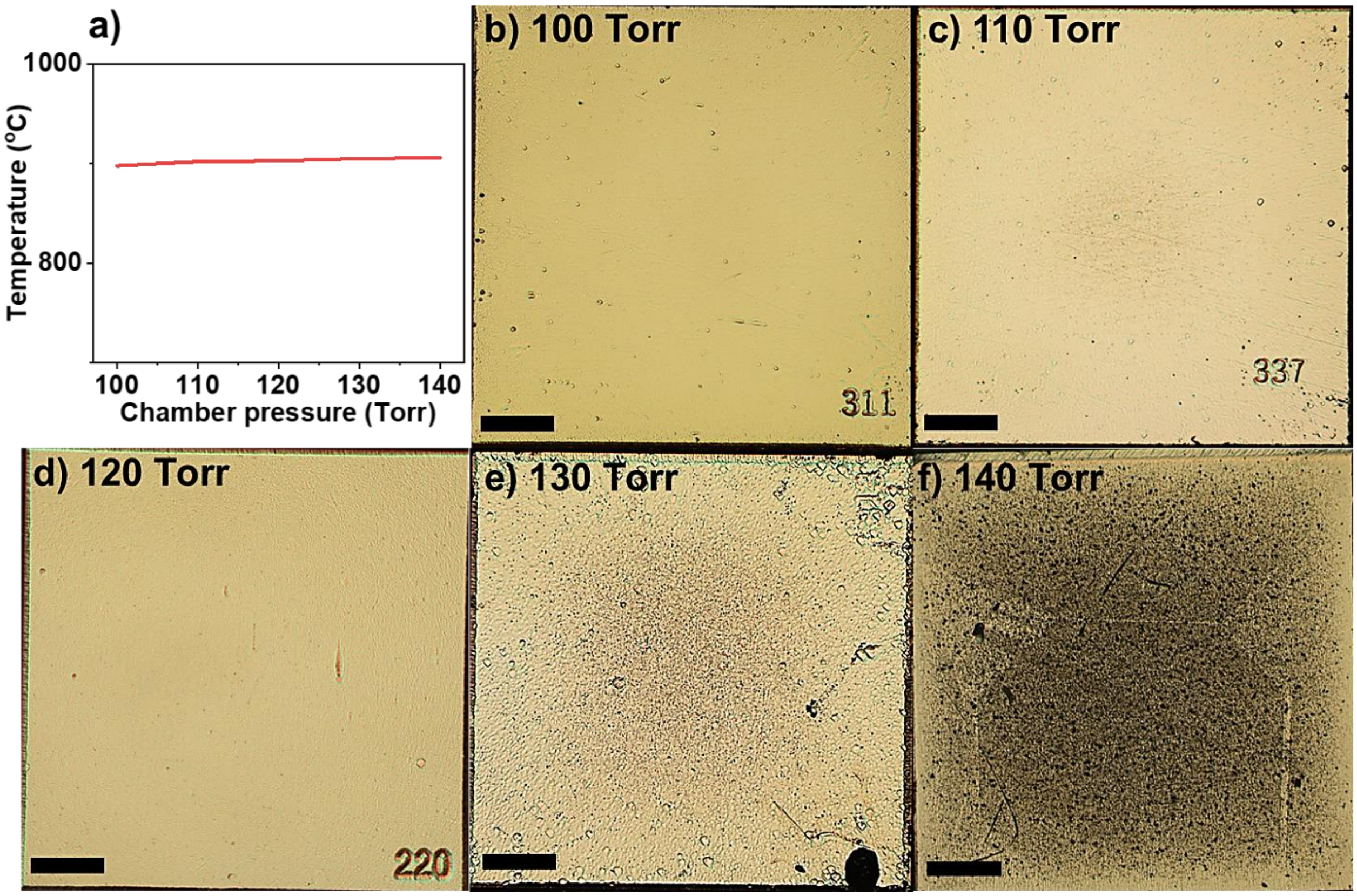

**Fig. 6.** (a) Diagram illustrating the sample surface temperature and chamber pressure relationship. (b-f) Optical images of the samples were deposited in different chamber pressure conditions ranging from 100 Torr to 140 Torr, respectively, at 2.5x magnification. The scale in all images is 500 µm.

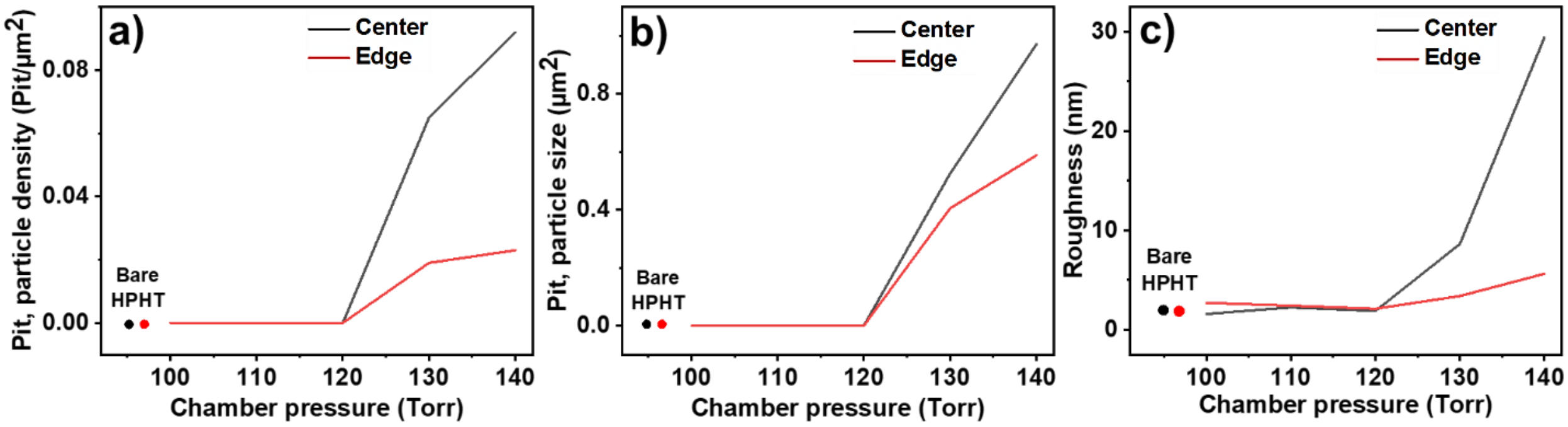

**Fig. 7.** The diagrams illustrate the relationships between a) the pit density, b) average pit size, and c) surface roughness at both the center and edge regions of the samples with the chamber pressure.

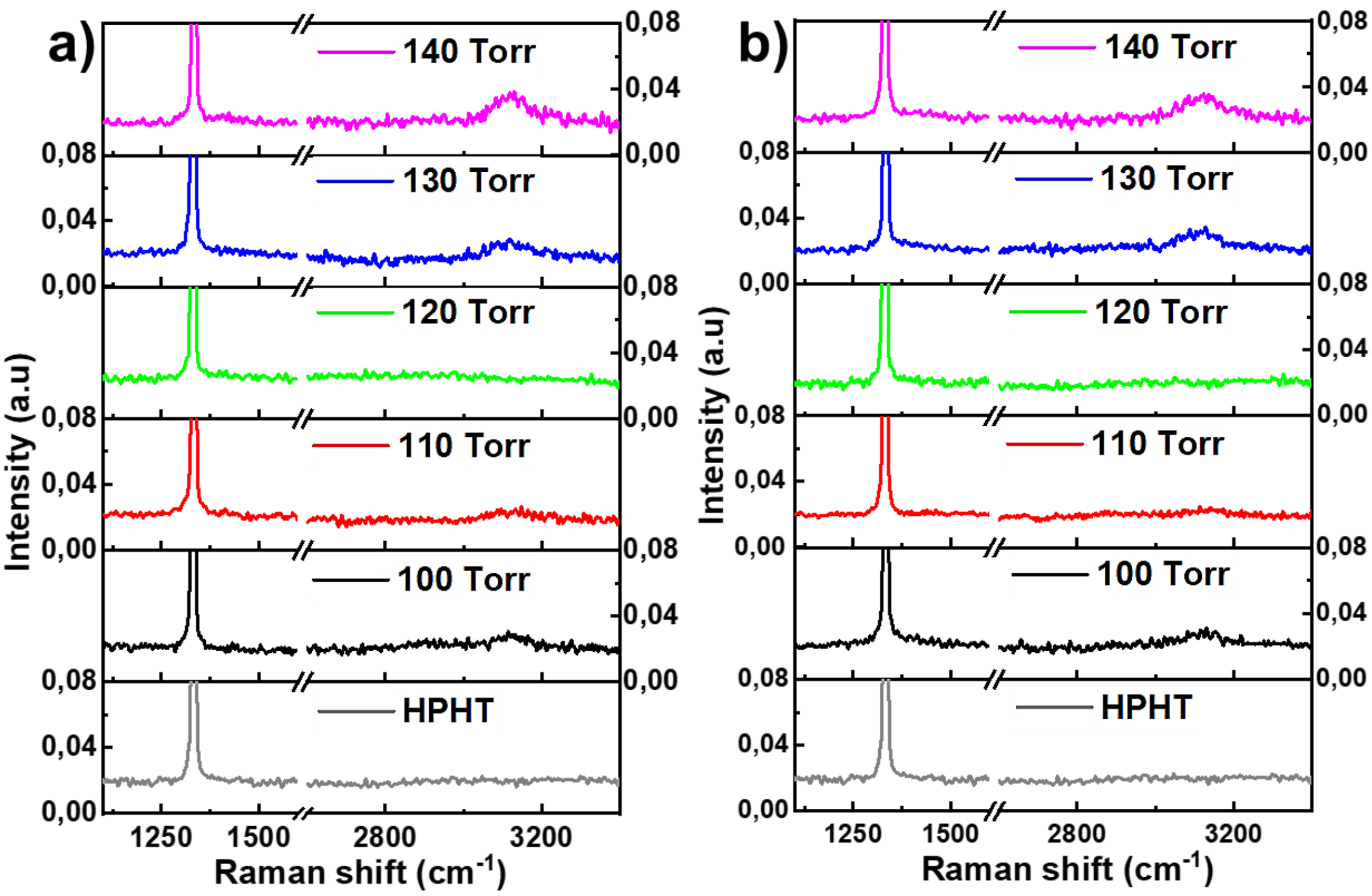

**Fig. 8.** Raman spectra at (a) the center and (b) the edge of samples deposited in different chamber pressures ranging from 100 Torr to 140 Torr.

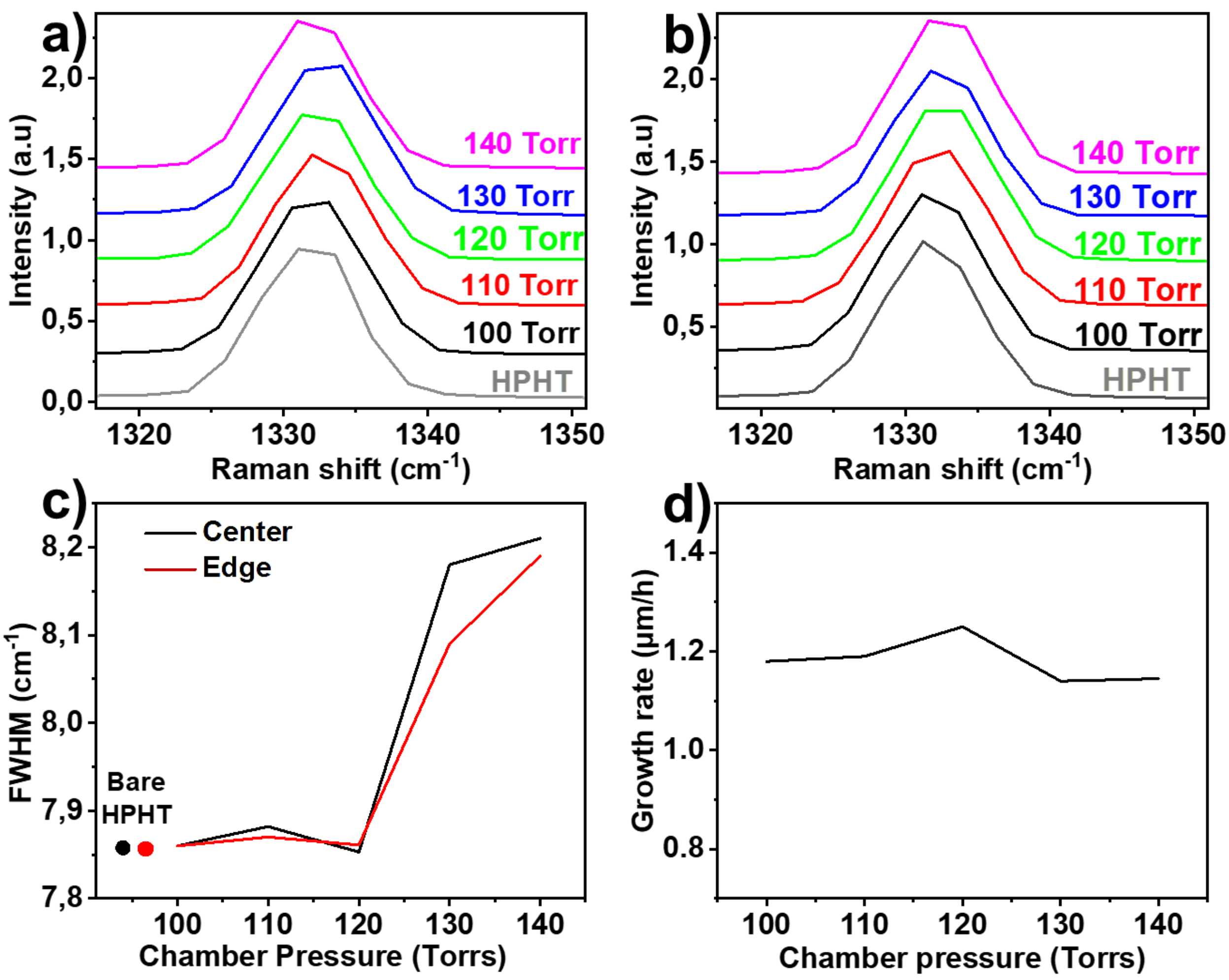

**Fig. 9.** The single diamond peak at (a) the center and (b) the edge of samples deposited in different conditions of chamber pressure, (c) the comparison of FWHM of samples, and (d) the diagram reveals the relationship between the growth rate and chamber pressure.

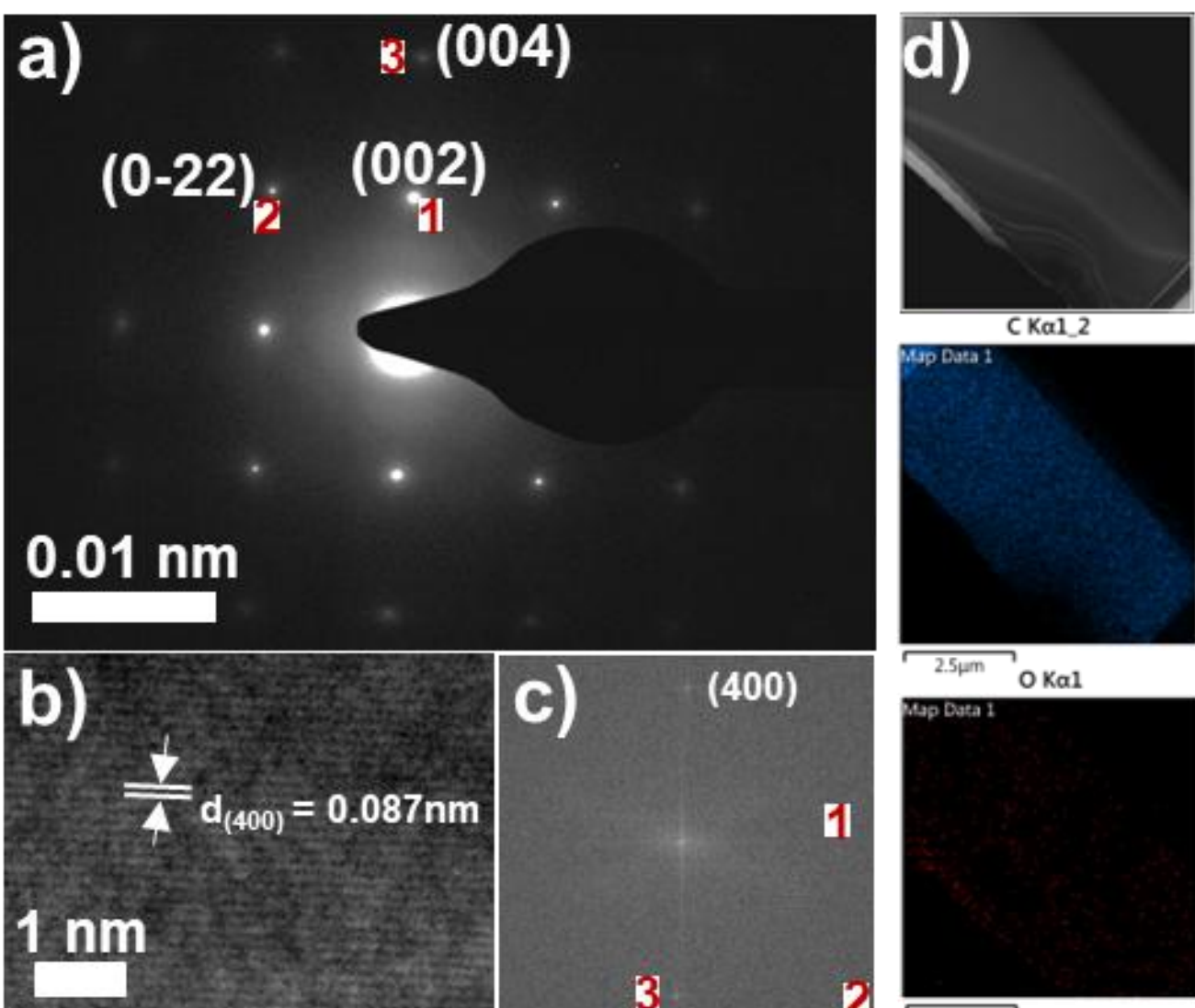

**Fig. 10.** a) The image of SAED patterns, b) the HRTEM images at c) the corresponding FFT of HRTEM and d EDS mapping at the center of the sample grown at 3900 W and 120 Torr.

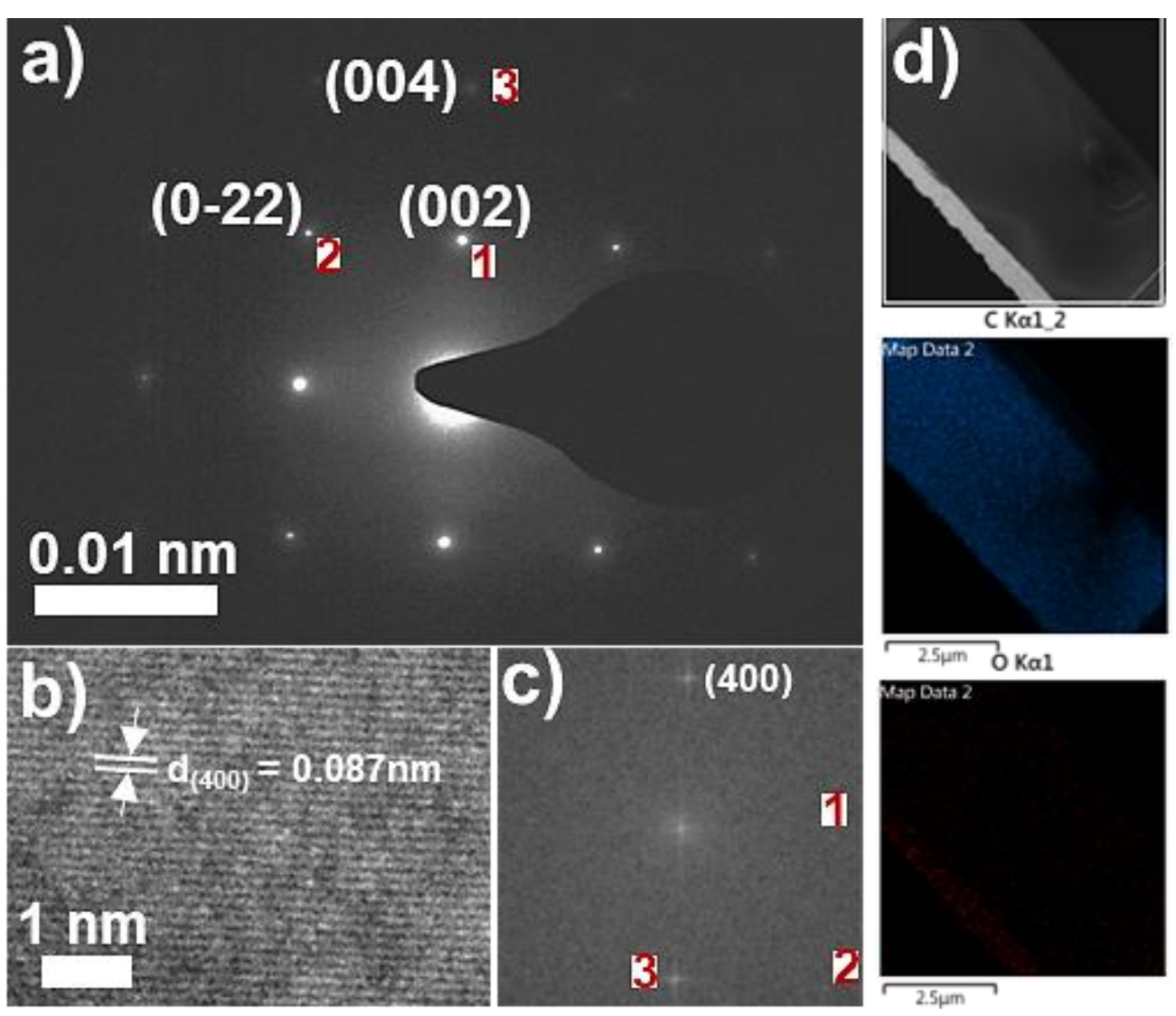

**Fig. 11.** a) The image of SAED patterns, b) the HRTEM images at c) the corresponding FFT of HRTEM and d EDS mapping at the edge of the sample grown at 3900 W and 120 Torr.